\documentclass[prd,aps,tightline,twocolumn,nofootinbib,superscriptaddress]{revtex4}

\usepackage{amssymb}
\usepackage{amsmath}
\usepackage{graphicx}
\usepackage{latexsym}

\begin{document}

%\hfill ICRR ??? \\
%\hfill \today

%\vskip 1.35cm

%\vspace{0.5cm}
\title{ Implications of CDMS II result on Higgs sector in the MSSM}
%\vspace{1cm}

\author{Junji Hisano}
\affiliation{Institute for Cosmic Ray Research, University of Tokyo, Kashiwa 277-8582, Japan}
\affiliation{Institute for the Physics and Mathematics of the Universe,
University of Tokyo, Kashiwa 277-8568, Japan}

\author{Kazunori Nakayama}
\affiliation{Institute for Cosmic Ray Research, University of Tokyo, Kashiwa 277-8582, Japan}

\author{Masato Yamanaka}
\affiliation{Institute for Cosmic Ray Research, University of Tokyo, Kashiwa 277-8582, Japan}

%\vskip 0.15in

%%%%%%%%%%%%%%%%%%%%%%%%%%%%%%%%%%%%%%%%%%%%%%%%%%%%%%%%
\begin{abstract}
	We study implications of two dark matter candidate events at 
	CDMS-II on the neutralino dark matter scenario in the supersymmetric standard model,
	in light of the recent lattice simulation on the strange quark content of a nucleon.
	The scattering rate of neutralino-nucleon is dominated by Higgs exchange processes and
	the mass of heavy Higgs boson is predicted for the neutralino of Bino-Higgsino mixing state.
	In the case of Wino-Higgsino mixing, the Higgs sector may not be constrained.	
\end{abstract}
%%%%%%%%%%%%%%%%%%%%%%%%%%%%%%%%%%%%%%%%%%%%%%%%%%%%%%%%

%\setcounter{footnote}{0}

%\tableofcontents
\maketitle

%%%%%%%%%%%%%%%%%%%%%%%%%%%%%%%%%%%%%%%%%%%%%%%%%%%%%
%\section{Introduction} %%%%%%%%%%%%%%%%%%%%%%%%%%%%%%%%%%%%%%%%%%
%%%%%%%%%%%%%%%%%%%%%%%%%%%%%%%%%%%%%%%%%%%%%%%%%%%%%

{\it Introduction :}
The existence of non-baryonic dark matter (DM) has been established by cosmological observations 
\cite{Dunkley:2008ie}. Weakly interacting massive particles (WIMPs) are attractive candidate, 
and, in particular, the lightest neutralino, 
$\tilde{\chi}^0$,  in the minimal supersymmetric standard model (MSSM) is the most extensively studied among them. It is a linear combination of superpartners 
of $U(1),~SU(2)$ gauge bosons and two neutral Higgs bosons (Bino $\tilde B$, Wino $\tilde W$, and 
Higgsinos $\tilde H_{1(2)}$) and is stable due to the R-parity conservation when it is the lightest 
supersymmetric particle (LSP). 
%Even though they are produced at
%the Large Hadron Collider (LHC), it would be difficult to derive its  mass and  mixing structure, 
%because they leave detectors with no signatures. 

One of the methods for probing the $\tilde{\chi}^0$ nature is the direct detection experiments 
of WIMP DM. In the experiments, one searches for the signatures of neutralino-nucleon 
($\tilde{\chi}^0$-$N$) scattering. The scattering rate has been calculated in various 
scenarios in the supersymmetric (SUSY) framework~\cite{Drees:1993bu,Jungman:1995df}, 
and is found to be sensitive to $\tilde{\chi}^0$ mass and its mixing matrix. 
%%%
On a parallel with theoretical arguments, many experiments have searched for the DM signals, 
and their sensitivities have been improved. The upper limit on the $\tilde{\chi}^0$-$N$ 
spin-independent (SI) cross section had been obtained by XENON10, $\sigma^{\rm SI} < 4.5 
\times 10^{-8}$pb for  a DM of mass 30~GeV~\cite{Angle:2007uj} and CDMS, $\sigma^{\rm SI} < 
4.6 \times 10^{-8}$pb for 60~GeV~\cite{Ahmed:2008eu}. 
%%%
Experiments also have imposed on the upper limit on the spin-dependent (SD) cross section. 
The upper limits had been obtained by XENON10 as 
$\sigma^{\rm SD} < 6.0 \times 10^{-3}$pb for a DM of mass  
30~GeV~\cite{Angle:2008we}. 
In addition, the SD cross section is constrained by searching for
neutrinos coming from DM annihilation in the Sun. 
For a neutralino mass lighter than 100~GeV, the most stringent bound comes from
Super-Kamiokande~\cite{Desai:2004pq}, and for higher mass
AMANDA~\cite{Ackermann:2005fr} and 
IceCube with 22 strings give the stringent limits~\cite{Abbasi:2009uz}.
%The most stringent limit is from 
%Super-Kamiokande, $\sigma^{SD} < ??? \times 10^{-3}$pb \cite{???}. 
%%%
These limits have already put constraints on the parameter region of the MSSM.

Very recently, the CDMS-II experiment reported the final 
results from the 5-Tower WIMP search~\cite{cdms:2009zw}.
They found two events in a signal region, which may be interpreted as 
signals from DM-induced nuclear recoils~\cite{Kadastik:2009gx}.
Although the confidence level is poor because of the expected  background of around
0.8 event, it is still worth studying the implications of new CDMS result
for the purpose of providing a possible direction of future DM searches.

In this letter, we extract the possible nature of detected WIMPs from the result of CDMS-II, 
and derive the expected parameter space of the MSSM. In particular, we focus on the cases 
that $\tilde{\chi}^0$ dominantly consists of $\tilde B$ and $\tilde H_{1(2)}$, and
that of $\tilde W$ and $\tilde H_{1(2)}$.
The relevant parameters for identifying the nature of $\tilde{\chi}^0$ are mainly two mass 
parameters, $M_1 (M_2)$ and $\mu$, and one dimensionless parameter, tan$\beta$. 
Here $M_1 (M_2)$ and $\mu$ are Bino (Wino) mass and Higgsino mass, respectively, 
and tan$\beta$ is the ratio between vacuum 
expectation values of up-type Higgs and down-type Higgs. 
When we discuss 
the $\tilde{\chi}^0$-$N$ SI cross section, heavy Higgs mass, $m_{H^0}$, is also one of the important parameters. 
 By varying the values of these parameters, we search for the parameter region that 
is consistent with both the CDMS-II result and WMAP result for the relic abundance of DM, and predict 
the heavy Higgs mass. 
%%%
In addition, by applying the analysed results on the physics of indirect detection of DM, we predict 
the limit on the its rate of neutrinos coming from DM annihilation in the sun.

%%%%%%%%%%%%%%%%%%%%%%%%%%%%%%%%%%%%%%%%%%%%%%%%%%%%%
%\section{Direct detection of neutralino dark matter} %%%%%%%%%%%%%%%%%%%%%%%%%%%%%%%
%%%%%%%%%%%%%%%%%%%%%%%%%%%%%%%%%%%%%%%%%%%%%%%%%%%%%

{\it Direct detection of neutralino dark matter : }
Before investigating the parameter space, it is instructive to recall the physics of direct detection 
of DM and the neutralino mass matrix. 

For direct detection, each $\tilde \chi^0$-$N$ scattering cross section includes two type 
contributions, Higgs and squark exchange for the SI interaction, and $Z$-boson and squark exchange 
for the SD interaction. The contribution of squark exchange is proportional to $m_{\tilde q}^{-4}$ 
and is typically subdominant, thus, in this letter, for simplicity we neglect them.   
%%%
The $\tilde \chi^0$-$N$ scattering cross section is given by~\cite{Jungman:1995df}
\begin{equation}
 \begin{split}
    \sigma = &\frac{4}{\pi} 
    \left( \frac{m_{\tilde \chi^0} m_T}{m_{\tilde \chi^0} + m_T} \right)^2 \\
    &\times \biggl[ \bigl( n_p f_p + n_n f_n \bigr)^2 
    + 4 \frac{J+1}{J} 
    \bigl( a_p \langle s_p \rangle + a_n \langle s_n \rangle \bigr)^2    \biggr], 
 \end{split}     \label{2a}
\end{equation}
where the first and the second term in the bracket are the contributions of SI and SD interaction, 
respectively.  $m_T$ is the mass of target nucleus. 
%%%
$n_p (n_n)$ is the number of proton (neutron) in the target nucleus, and $f_{p}$ is given by 
\begin{equation}
 \begin{split}
    f_{p} 
    =& \sum f_q^{H} \langle p  | \bar q q | p  \rangle \\
    =& \sum_{q=u, d, s} \frac{f_q^{H}}{m_q} m_p f_{T_q}^{(p)} 
    + \frac{2}{27}f_{T_G} \sum_{q=c, b, t} \frac{f_q^{H}}{m_q} m_p.
 \end{split}     \label{2b}
\end{equation}
where $f_{T_G}=1-\sum_{u,d,s}f_{T_q}^{(p)}$.
The second term comes from coupling of heavy quarks to gluons through trace anomalies
\cite{Shifman:1978zn}. 
$f_n$ is derived from Eq.~(\ref{2b}) with exchange $p\leftrightarrow n$. 
Here $m_p$ ($m_n$) stands for the proton (neutron) mass. For 
the nucleon mass matrix elements, we take $f_{T_u}^{(p)} = 0.023$, $f_{T_d}^{(p)} = 0.034$, 
$f_{T_u}^{(n)} = 0.019$, $f_{T_d}^{(n)} = 0.041$
\cite{Gasser:1990ce, Adams:1995ufa}
and $f_{T_s}^{(p)} = f_{T_s}^{(n)} = 0.025$~\cite{Ohki:2008ff}.  Notice that the strange quark content of 
the nucleon $f_{T_s}$ is much smaller than previously thought according to the recent 
lattice simulation~\cite{Ohki:2008ff}, and this leads to a significant suppression on the SI cross section for $\tilde \chi^0-N$ scattering. 
The effective coupling between the neutralino 
and nucleon through the Higgs exchange, $f_q^{H}$, is given by 
\begin{equation}
 \begin{split}
    f_q^{H} = m_q \frac{g_2^2}{4 m_W} 
    \biggl( \frac{C_{h \tilde \chi \tilde \chi}  C_{hqq}}{m_{h^0}^2} 
    + \frac{C_{H \tilde \chi \tilde \chi}  C_{Hqq}}{m_{H^0}^2}     \biggr), 
 \end{split}     \label{2c}
\end{equation}
where $h^0$ and $H^0$ are the SM-like Higgs and heavy Higgs, respectively, 
$C_{h \tilde \chi \tilde \chi}$ ($C_{H \tilde \chi \tilde \chi}$) stands for the 
$h^0$($H^0$)-$\tilde \chi^0$-$\tilde \chi^0$ coupling, $C_{hqq}$ ($C_{Hqq}$) is the 
$h^0$($H^0$)-quark-quark Yukawa coupling, explicit expressions of them are given in 
literatures~\cite{Bergstrom:1995cz, Drees:1992rr, Drees:1993bu,Jungman:1995df}. 
%%%
$J$ is the total nuclear spin, $a_p$ and $a_n$ are the effective $\tilde \chi^0$-$N$ couplings, 
and $\langle s_{p(n)} \rangle = \langle N | s_{p(n)} |N \rangle$ are the expectation values 
of the spin content of the proton and neutron groups within the nucleus. Detailed nuclear 
calculations for $\langle s_{p(n)} \rangle$ exist in literature \cite{Engel:1989ix}. 

Before going on, let us see in which situation the scattering cross section becomes large enough 
to be detected at direct detection experiments.
In the gauge-eigenstate basis $(\tilde B, \tilde W, \tilde H_1, \tilde H_2)$, 
the neutralino mass matrix  $\mathcal{M}_N$ is given as  
\[ 
%\mathcal{M} = 
\left(
  \begin{array}{cccc}
   M_1 & 0 & - m_Z s_W c_\beta  & m_Z s_W s_\beta  \\
   0  & M_2 & m_Z c_W c_\beta & - m_Z c_W s_\beta  \\
   - m_Z s_W c_\beta & m_Z c_W c_\beta & 0 & - \mu  \\
   m_Z s_W s_\beta & - m_Z c_W s_\beta & - \mu & 0
  \end{array}
\right),  \] 
where $M_1, M_2$ and $m_Z$ are masses of Bino, Wino and $Z$-boson, respectively, and we have 
introduced abbreviations $s_W = \sin \theta_W$, $c_W = \cos \theta_W$, $t_W = 
\tan \theta_W$, $c_\beta = \cos \beta$ and $s_\beta = \sin \beta$. 
%%%
$h^0$($H^0$)-$\tilde \chi^0$-$\tilde \chi^0$ couplings in Eq. (\ref{2c}) are yielded 
through the mixing of Bino (Wino) component with Higgsino component in the 
diagonalizing 
matrix of neutralino. Therefore, when the mixing angle is large, {\it i.e.},
$M_1 (M_2) \simeq \mu$, the direct detection rate is enhanced.

Here we show the qualitative behavior of $h^0$($H^0$)-$\tilde \chi^0$-$\tilde \chi^0$ coupling
in the limit case of $m_{H^0} = m_A$, where $m_A$ stands for the mass of CP-odd Higgs boson.
For the Bino-like $\tilde{\chi}^0$ ($M_1 \ll M_2, \mu$), the diagonalizing matrix 
of neutralino is calculated perturbatively, and $C_{h \tilde \chi \tilde \chi}$ and 
$C_{H \tilde \chi \tilde \chi}$ are approximated as follows 
\begin{equation}
 \begin{split}
    &C_{h \tilde \chi \tilde \chi} \simeq \frac{m_Z s_W t_W}{M_1^2 - \mu^2}
    \bigl[ M_1 + \mu \sin2 \beta \bigr], \\
    &C_{H \tilde \chi \tilde \chi} \simeq - \frac{m_Z s_W t_W}{M_1^2 - \mu^2}
    \mu \cos2\beta .
 \end{split}     \label{2d}
\end{equation}
Notice that this perturbative calculation breaks down if $|M_1 - |\mu|| \lesssim 
m_Z$. Similarly, they are calculated as follows for the Wino-like $\tilde{\chi}^0$ ($M_2 \ll M_1, |\mu|$), 
\begin{equation}
 \begin{split}
    &C_{h \tilde \chi \tilde \chi} \simeq \frac{m_Z c_W}{M_2^2 - \mu^2}
    \bigl[ M_2 + \mu \sin2 \beta \bigr], \\
    &C_{H \tilde \chi \tilde \chi} \simeq - \frac{m_Z c_W}{M_2^2 - \mu^2}
    \mu \cos2\beta , 
 \end{split}     \label{2e}
\end{equation}
and for the Higgsino-like $\tilde{\chi}^0$ ($|\mu| \ll M_1, M_2$), 
\begin{equation}
 \begin{split}
    &C_{h \tilde \chi \tilde \chi} \simeq \frac{1}{2} \bigl[ 1 \pm \sin2\beta \bigr]
    \biggl[ t_W^2 \frac{m_Z c_W}{M_1 - |\mu|}  + \frac{m_Z c_W}{M_2 - |\mu|} \biggr], \\
    &C_{H \tilde \chi \tilde \chi} \simeq \pm \frac{1}{2} \cos2\beta 
    \biggl[ t_W^2 \frac{m_Z c_W}{M_1 - |\mu|}  + \frac{m_Z c_W}{M_2 - |\mu|} \biggr] .
 \end{split}     \label{2f}
\end{equation}
In every case, since couplings are suppressed by SUSY mass parameters $M_1$, $M_2$, and 
$\mu$, smaller value of them leads the enhancement of direct detection rate of $\tilde{\chi}^0$ 
DM. The $\tilde{\chi}^0$-$N$ SI cross section also depends upon $\tan\beta$. In the limit 
case of $m_{H^0} = m_A$, Yukawa coupling for down-type quarks, $C_{Hdd}$, are proportional to 
$\tan\beta$. When $\tan\beta$ is large, therefore, the contribution of heavy Higgs boson 
becomes dominant.

%%%%%%%%%%%%%%%%%%%%%%%%%%%%%%%%%%%%%%%%%%%%%%%%%%%%%
%\section{Numerical result} %%%%%%%%%%%%%%%%%%%%%%%%%%%%%%
%%%%%%%%%%%%%%%%%%%%%%%%%%%%%%%%%%%%%%%%%%%%%%%%%%%%%

{\it Numerical result :}
The dominant contribution to the SI cross section
comes from light neutral Higgs boson exchange,
but generically this contribution is not enough to explain the observed CDMS events.
Hence the heavy Higgs exchange contribution must be added 
with an appropriate magnitude.

\begin{figure}
 \begin{center}
   \includegraphics[width=0.9\linewidth]{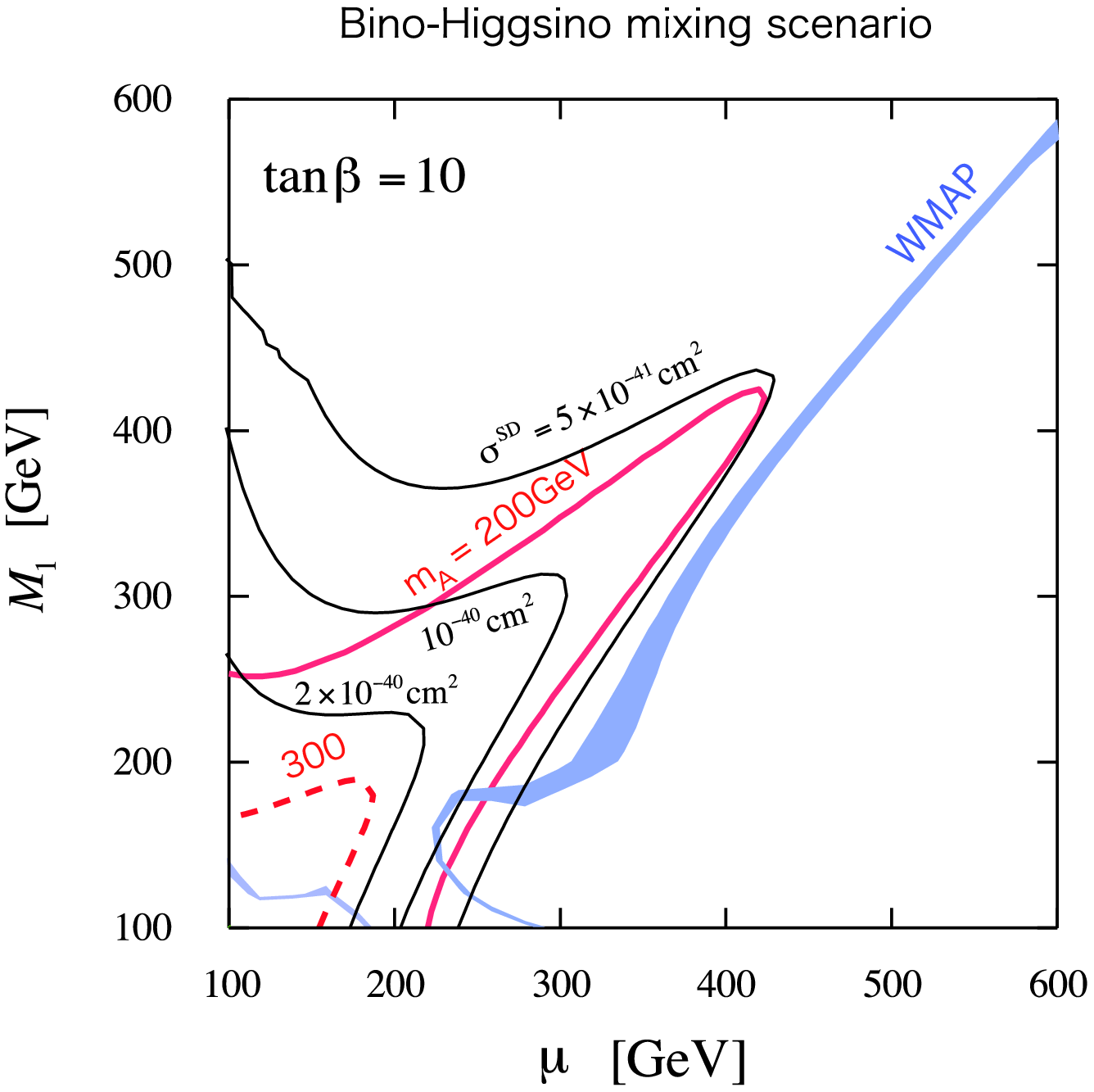} 
   %\vskip 1cm
   \includegraphics[width=0.9\linewidth]{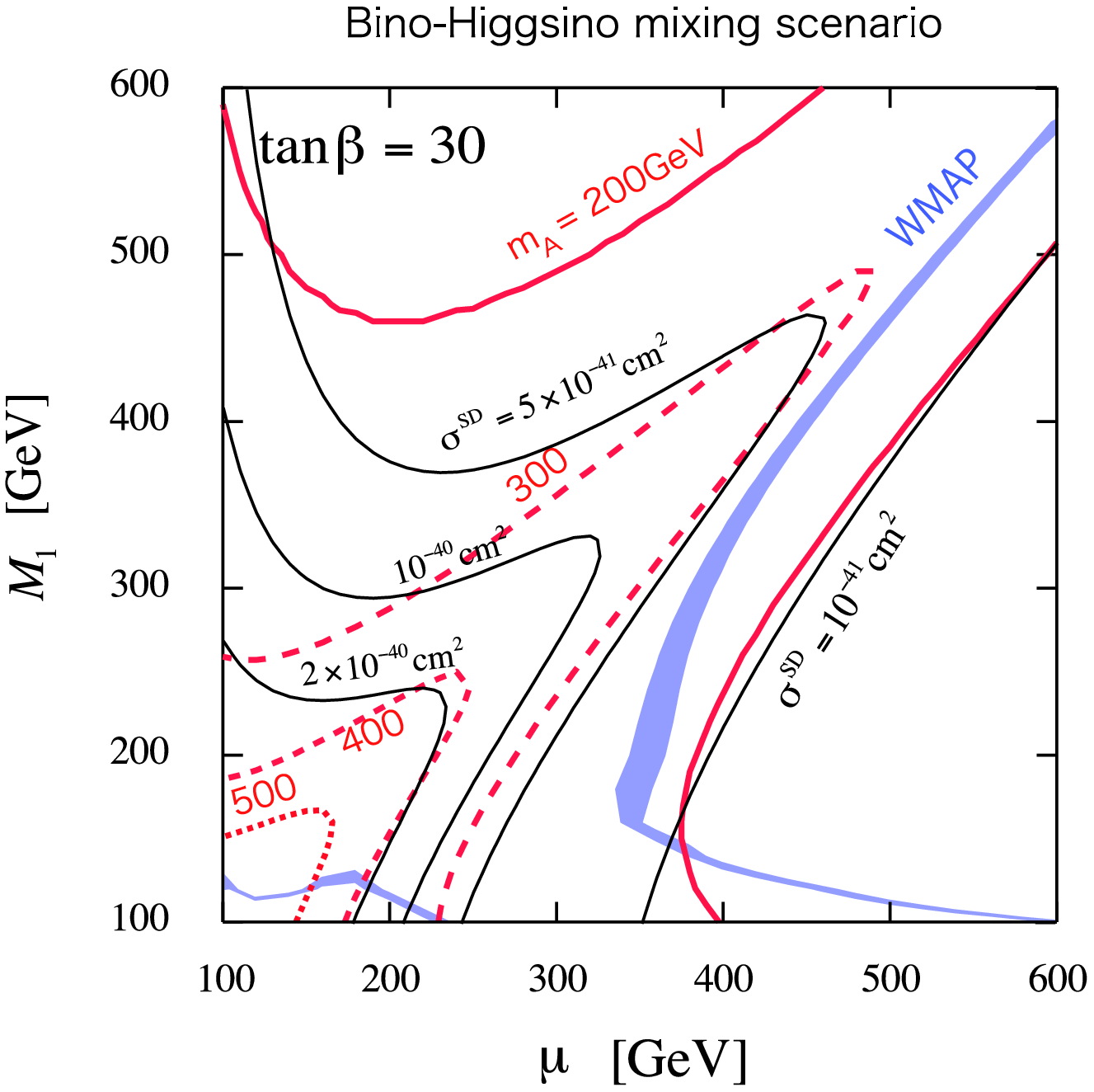} 
   \caption{ 
   	Contours of $m_A(=200, 300, 400, 500~{\rm GeV})$ on $(\mu,M_1)$-plane
	for reproducing the CDMSII events, for
	$\tan \beta=10$ (top) and $\tan \beta=30$ (bottom).
	The blue band shows the region where the relic abundance of neutralino is
	consistent with the WMAP result.
	Contours of SD cross section are also shown for estimating the 
	muon flux from the Sun.
   }
   \label{fig:B}
 \end{center}
\end{figure}

\begin{figure}
 \begin{center}
   \includegraphics[width=0.9\linewidth]{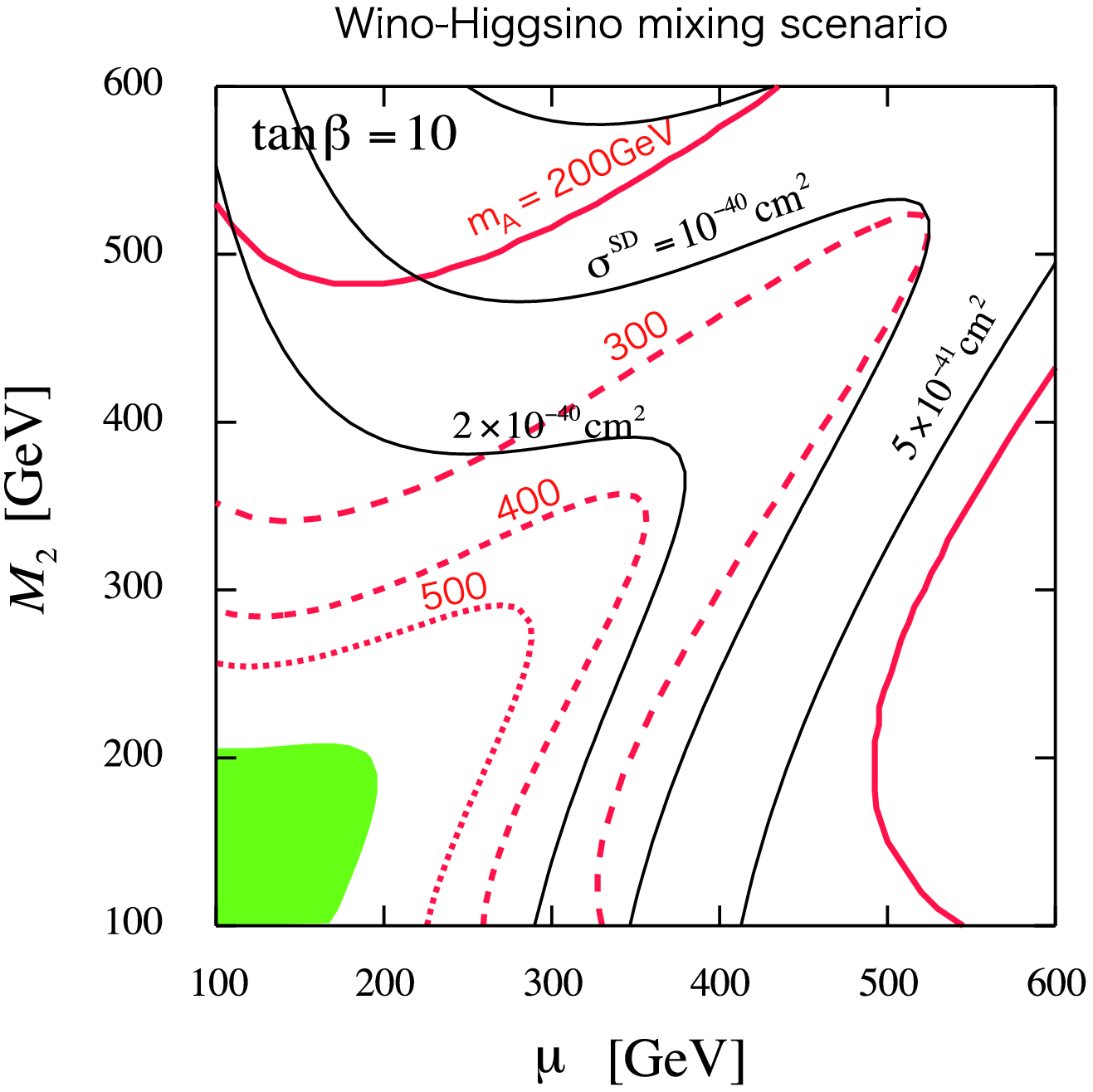} 
   %\vskip 1cm
   \includegraphics[width=0.9\linewidth]{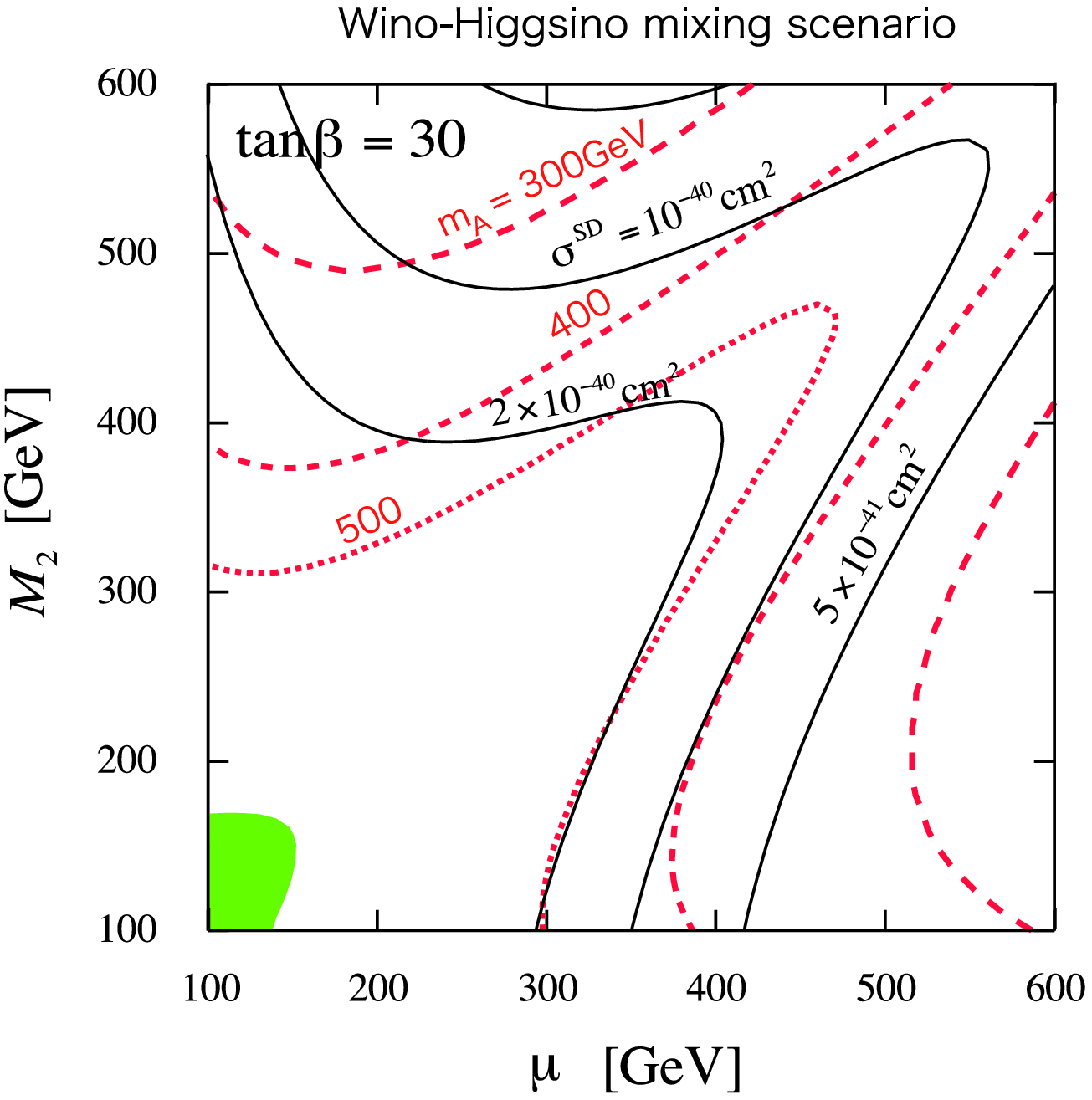} 
   \caption{ 
   	Same as Fig.~\ref{fig:B} but for $(\mu, M_2)$-plane.
	The green region at the bottom-left of the figure predicts
	too large SI cross section without including the heavy-Higgs contribution.
    }
   \label{fig:W}
 \end{center}
\end{figure}

Numerical results are shown in Fig.~\ref{fig:B} for the Bino-Higgsino mixing case
and in Fig.~\ref{fig:W} for the Wino-Higgsino mixing case.
In the Bino (Wino)-Higgsino mixing case, the Wino (Bino) mass $M_2$ ($M_1$)
is set to be sufficiently heavy.
Contours of the pseudo-scalar Higgs mass $m_A (=200, 300, 400, 500~{\rm GeV})$ 
for reproducing the CDMS-II events are shown on a $\mu$$-$$M_1$ plane in Fig.~\ref{fig:B}
and a $\mu$$-$$M_2$ plane in Fig.~\ref{fig:W}. 
Here, we demanded $\sigma^{\rm SI}/m_{\tilde \chi_0} = 3\times 10^{-46}~{\rm cm^2/GeV}$
for $m_{\tilde \chi_0} \gtrsim 100~{\rm GeV}$ as a typical relation 
consistent with the CDMS events.
We fix the light Higgs mass as $m_h=115~$GeV.
%The value of $m_A$ is not acceptable in the gray shaded regions.
It is seen that the typical mass of heavy Higgs boson must be rather light.
This explicitly shows that the heavy Higgs exchange contribution is important.
Only in the Wino-Higgsino mixing case, it is possible that the only light Higgs contribution
can explain the CDMS-II result, if the neutralino mass is light enough
as indicated by green-shaded region in Fig.~\ref{fig:W}.

In the Bino-Higgsino mixing case, we have also calculated the relic abundance of the neutralino
under the standard thermal freeze-out scenario using the DarkSUSY code~\cite{Gondolo:2004sc}
and shown the parameter region consistent with the WMAP result~\cite{Dunkley:2008ie}.
In this calculation we have set all the squark and slepton masses are heavy (=2~TeV)
so that the coannihilation between the neutralino and squarks/sleptons do not work.
Since the heavy Higgs bosons are light, the enhancement of the
neutralino annihilation cross section through $S$-channel resonance is available
as seen in the figure.
In the case of Wino-Higgsino mixing, there is no appropriate parameter regions
which fit the WMAP result.\footnote{
However, the Wino-like LSP is often realized in the anomaly-mediated SUSY breaking models
\cite{Randall:1998uk} where the gravitino is heavy enough to decay well before
big-bang nucleosynthesis (BBN) begins,
and the decay of gravitino can produce LSPs,
without disturbing the success of BBN~\cite{Kawasaki:2008qe}.
The resulting LSP abundance falls into a correct range favored by WMAP
depending on the reheating temperature after inflation~\cite{Bolz:2000fu}.
}

In order to discuss the detectability of the neutralino LSP at neutrino detectors, we have shown the
contours of SD cross section between a neutralino and proton,
which is related to a muon flux from the Sun
~\cite{Griest:1986yu,Ritz:1987mh,Kamionkowski:1991nj,Jungman:1994jr}.
Since the neutralinos scatters off nucleons in the Sun and then trapped inside the Sun,
the number density of neutralinos is significantly enhanced at the interior of the Sun~\cite{Spergel:1984re,Gould:1987ir}.
The total neutralino annihilation rate in the Sun is proportional to 
the neutralino-nucleon scattering cross section rather than its self-annihilation cross section,
because the number density is dynamically adjusted so that annihilation rate 
balances with the trapping rate, which is proportional to the scattering cross section.
High-energy neutrinos produced by DM annihilation in the Sun
can be detected as a muon signal at the neutrino detectors such as IceCube.
The scattering cross section required for the muon flux from the Sun is dominated by the 
SD one between neutralino and hydrogen atom through $Z$-boson exchange.
Since it also depends on the gaugino-Higgsino mixing, we have a definite prediction
on the resulting muon flux from the Sun for each parameter space.
As is seen from figures, SD cross section of $\mathcal O(10^{-41}-10^{-40})~{\rm cm^2}$
is predicted for large parameter space.
This may reach the sensitivity of the IceCube DeepCore experiment for 
these mass ranges~\cite{DeYoung:2009uk}.

%%%%%%%%%%%%%%%%%%FIGURE%%%%%%%%%%%%%%%%%%%

%%%%%%%%%%%%%%%%%%%%%%%%%%%%%%%%%%%%%%%%%

%%%%%%%%%%%%%%%%%%FIGURE%%%%%%%%%%%%%%%%%%%

%%%%%%%%%%%%%%%%%%%%%%%%%%%%%%%%%%%%%%%%%

%%%%%%%%%%%%%%%%%%%%%%%%%%%%%%%%%%%%%%%%%%%%%%%%%%%
%\section{Conclusions and Discussion} %%%%%%%%%%%%%%%%%
%%%%%%%%%%%%%%%%%%%%%%%%%%%%%%%%%%%%%%%%%%%%%%%%%%%

{\it Discussion :}
We have studied implications of the observed DM-like events at the CDMS-II detector
assuming that it is caused by the MSSM neutralino.
It is found that the heavy-Higgs contribution is necessary 
%in order to explain the observed events 
for the Bino-Higgsino mixing case,
taking into account the recent lattice simulation of the strange quark content in a nucleon.
In the case of Wino-Higgsino mixing, 
heavy-Higgs contributions are not always necessary.
In both of them, the mixing between gaugino and Higgsino must be large enough.

Some comments are in order.
Notice that relatively light Higgs bosons may be favored from the
viewpoint of fine-tuning issues (see e.g.,
Ref.~\cite{Kitano:2005ew}). However, 
there is a danger of obtaining too large $b\to s\gamma$ branching ratio
due to the charged Higgs loop contribution.
This contribution must be compensated by the destructive contribution 
from chargino-squark loops in order not to be contradict with observations.
This is indeed possible for sizable squark masses and $A$-terms. 

In addition, we should also pay attention to the additional
contributions of additional Higgs bosons to decays of pseudoscalar
mesons, such as $B^\pm \rightarrow \tau^\pm \nu$ \cite{Hou:1992sy},
$B_s \rightarrow \mu^+ \mu^-$ \cite{Babu:1999hn}, and $D_s^\pm
\rightarrow \tau^\pm (\mu^\pm) \nu$ \cite{Akeroyd:2009tn}.  
%Now there
%are 2-$\sigma$ level discrepancies between the SM predictions and the
%observations in $B^\pm \rightarrow \tau^\pm \nu$ and $D_s^\pm
%\rightarrow \tau^\pm (\mu^\pm) \nu$, and the tension is increased when
%considering the charged Higgs contributions the them. 
%
Connecting them
to the prospective constraint on parameter space with the improvement
of direct detection experiments, it would be possible to more tightly
constrained the masses of heavy Higgs bosons. 

When the gaugino-Higgsino mixing is large, 
the studies of SUSY events at the LHC would be more fruitful. Even in a case that Bino and/or Wino are lighter than Higgsinos, the heavier neutralinos and chargino have sizable gaugino components. In the case, the cascade decay of squarks produces the heavier neutralinos and chargino so that we could measure all of the parameters in the chargino and neutralino mass matrices~\cite{Hisano:2008ng}. 

We also comment on a possible relation to the positron excess observed by the PAMELA
satellite~\cite{Adriani:2008zr}.
Generically the neutralino LSP annihilates hadronically and 
it is difficult to fit the positron spectrum.
However, in the case of Wino-like neutralino LSP, the main annihilation mode is
into $W^+W^-$ and the subsequent decay of $W$ into $e\nu$ may explain the
positron excess for appropriate diffusion models of the galaxy 
with anti-proton bound marginally satisfied~\cite{Hisano:2008ti}.Therefore the light Wino can both explain the CDMS-II and PAMELA result,
if the Wino-Higgsino mixing is large enough.
For the case of Higgsino-like LSP, the situation is similar, but 
an order of magnitude larger boost factor than the Wino case is required
for reproducing the PAMELA positron excess.

Although the statistical significance of DM candidate events at the CDMS-II detector is poor,
it can soon be checked by forthcoming XENON100 and/or XMASS experiments.
If similar signals will be found in those experiments, it definitely has implications on the
Higgs sector as studied in this letter.

%%%%%%%%%%%%%%%%%%%%%%%%%%%%%%%%%%%%%%%%%%%%%%%%%%
\section*{Acknowledgments} %%%%%%%%%%%%%%%%%%%%%%%%%%%%%%%%%%%%
%%%%%%%%%%%%%%%%%%%%%%%%%%%%%%%%%%%%%%%%%%%%%%%%%%
K.N. would like to thank the Japan Society for the Promotion of
Science for financial support.  The work was supported in part by the
Grant-in-Aid for the Ministry of Education, Culture, Sports, Science,
and Technology, Government of Japan, No. 20244037 and No. 2054252
(J.H.) and No. 20007555 (M.Y.).

%\clearpage%%%%%%%%%%%%%%%%%%%%%%%%%%%%%

\end{document}